\documentclass[prl,letterpaper,twocolumn,showpacs,showkeys,lengthcheck,tightenlines,
               nofootinbib,preprintnumbers,superscriptaddress]{revtex4-1} 

\usepackage{mathtools}
\usepackage{amssymb,amsfonts}
\usepackage{bm}
\usepackage{slashed}

\usepackage[svgnames]{xcolor}
\usepackage[english]{babel}
\usepackage{blindtext}
\usepackage{microtype}
\usepackage{tikz}
\usepackage{dcolumn}
\usepackage{multirow}
\usepackage[]{units}

\usepackage{graphicx}
\usepackage{float}
\usepackage[caption=false]{subfig}

\usepackage[colorlinks=true,urlcolor=blue]{hyperref}
\usepackage{ifpdf}

\graphicspath{{.}{figures/}}

\def\d{\delta}
\def\g{\gamma}

\def\ve{\varepsilon}

\def\l{\lambda}
\def\s{\sigma}
\def\o{\omega}

\def\O{\Omega}

\def\hs{\hspace}

\def\no{\nonumber}

\def\lf{\left}
\def\rg{\right}

\newcommand{\vect}[1]{\boldsymbol{#1}}

\newcommand{\sh}[1]{\slashed{#1}}

\font\bb=bbmss10 scaled 1200
\def\ident{\mbox{\bb 1}}

\begin{document}

\preprint{ADP-15-19/T921}

\title{Relativistic and Nuclear Medium Effects on the Coulomb Sum Rule}

\author{Ian~C.~Clo\"et}
\affiliation{Physics Division, Argonne National Laboratory, Argonne, Illinois 60439, USA}

\author{Wolfgang~Bentz}
\affiliation{Department of Physics, School of Science, Tokai University,
             Hiratsuka-shi, Kanagawa 259-1292, Japan}

\author{Anthony~W.~Thomas}
\affiliation{CSSM and ARC Centre of Excellence for Particle Physics at the Terascale, \\ 
             Department of Physics, University of Adelaide, Adelaide SA 5005, Australia}

\begin{abstract}
In light of the forthcoming high precision quasielastic electron scattering data from Jefferson Lab, it is timely for the various approaches to nuclear structure to make robust predictions for the associated response functions. With this in mind, we focus here on the longitudinal response function and the corresponding Coulomb sum rule for isospin-symmetric nuclear matter at various baryon densities. Using a quantum field-theoretic quark-level approach which preserves the symmetries of quantum chromodynamics, as well as exhibiting dynamical chiral symmetry breaking and quark confinement, we find a dramatic quenching of the Coulomb sum rule for momentum transfers $\lf|\vect{q}\rg| \gtrsim 0.5\,$GeV. The main driver of this effect lies in changes to the proton Dirac form factor induced by the nuclear medium. Such a dramatic quenching of the Coulomb sum rule was not seen in a recent quantum Monte Carlo calculation for carbon, suggesting that the Jefferson Lab data may well shed new light on the explicit role of QCD in nuclei. 
\end{abstract}


\maketitle
\looseness=-1
Traditionally the nucleus is viewed as a collection of nucleons that interact via phenomenological potentials. This picture has proven successful since the establishment of the nuclear shell model and the interim has seen steady refinement, culminating today in sophisticated non-relativistic quantum-many-body approaches~\cite{Wiringa:1994wb,Navratil:2009ut,Barrett:2013nh,Lovato:2013cua,Carlson:2014vla}. 
A key assumption of such approaches is that the internal structural properties of the 
nucleons which comprise a nucleus are the same as those of free nucleons. However, with the realization that quantum chromodynamics (QCD) is the fundamental theory of the strong interaction, it is natural to expect that these nucleon properties are modified by the nuclear medium~\cite{Guichon:1987jp,Guichon:1995ue,Saito:2005rv,Bentz:2001vc}. Understanding the validity of these two viewpoints remains a key challenge for contemporary nuclear physics. Should it turn out that nucleon properties are significantly modified by the nuclear medium, this would represent a new paradigm for nuclear physics and help build a bridge between QCD and nuclei. On the other hand, if the bound nucleon is unchanged this would shed light on colour confinement in QCD and force a rethink of numerous approaches to hadron structure.

Experimental evidence for explicit quark and gluon effects in nuclei remains elusive and to date the most famous example of such evidence -- albeit not incontrovertible -- is the EMC effect~\cite{Geesaman:1995yd,Norton:2003cb,Malace:2014uea}. First observed in 1982~\cite{Aubert:1983xm}, the EMC effect refers to a quenching of the nuclear structure functions relative to those of a free nucleon, and demonstrates that valence quarks in a nucleus carry a smaller momentum fraction than those in a free nucleon. Numerous explanations have been proposed, ranging from nuclear structure~\cite{Akulinichev:1990su,Dunne:1985ks,Dunne:1985cn,Bickerstaff:1989ch} to QCD effects~\cite{Clark:1985qu,Bickerstaff:1985ax,Saito:1992rm,Mineo:2003vc,Smith:2003hu,Cloet:2005rt}, however, no consensus has yet been reached concerning the cause of the EMC effect.

Although the EMC effect is best known, the first hints of QCD effects in nuclei came from quasielastic electron scattering on nuclear targets~\cite{Altemus:1980wt,Noble:1980my,Meziani:1984is}. The differential cross-section for this process -- with energy transfer $\omega$ and 3-momentum transfer $\lf|\vect{q}\rg|$ -- has the form~\cite{Wehrberger:1993zu}
\begin{multline}
\frac{d^2\sigma}{d\Omega\,d\omega} = \sigma_{\text{Mott}}
\biggl[\frac{q^4}{\lf|\vect{q}\rg|^4}\,R_L(\o,\lf|\vect{q}\rg|) \allowdisplaybreaks \\
+ \biggl(\frac{q^2}{2\lf|\vect{q}\rg|^2} + \tan^2\frac{\theta}{2}\biggr)\,R_T(\o,\lf|\vect{q}\rg|) \biggr],
\end{multline}
where $q^2 = -Q^2 = \omega^2 - \lf|\vect{q}\rg|^2$ and $\theta$ is the electron scattering angle. The first experiment to separately determine the longitudinal ($R_L$) and transverse ($R_T$) response functions was performed at MIT Bates on $^{56}$Fe~\cite{Altemus:1980wt}. The results suggested a significant quenching of the Coulomb sum rule (CSR)~\cite{McVoy:1962zz}:
\begin{align}
S_L(\lf|\vect{q}\rg|) &= \int_{\omega^+}^{\lf|\vect{q}\rg|}d\omega\ 
\frac{R_L(\o,\lf|\vect{q}\rg|)}{Z\,G_{Ep}^2(Q^2) + N\,G_{En}^2(Q^2)},
\label{eq:csr}
\end{align}
compared to the \textit{non-relativistic} expectation that $S_L(\lf|\vect{q}\rg|)$ should approach unity for $\lf|\vect{q}\rg|$ much greater than the Fermi momentum; on the proviso that the nucleon form factors are not modified by the nuclear medium~\cite{Noble:1980my} ($G_{Ep}$ and $G_{En}$ in Eq.~\eqref{eq:csr} are the free nucleon Sachs electric form factors and $\omega^+$ excludes the elastic peak).\footnote{Because the timelike region is not accessible in elastic scattering we define the CSR empirically by Eq.~\eqref{eq:csr}.}
The Bates result was soon verified at Saclay for various nuclei~\cite{Meziani:1984is}. 

Since that time the experimental situation has been in a state of flux. For example, a reanalysis of $^{12}$C, $^{40}$Ca and $^{56}$Fe world data performed in Refs.~\cite{Jourdan:1995np,Jourdan:1996np} -- utilizing an alternative prescription for the Coulomb corrections -- concluded that there is no evidence for quenching of the CSR. The analysis of the Coulomb corrections in those works was later challenged in Refs.~\cite{Aste:2005wc,Aste:2007sa,Wallace:2008ev} and quenching of the CSR over a wider range of nuclei was reported in Ref.~\cite{Morgenstern:2001jt}. This situation stands to be clarified however, as quasielastic electron scattering has been revisited in a comprehensive set of measurements at Jefferson Lab~\cite{E05110}. 

With this in mind, in this Letter we present predictions for the longitudinal response function and the CSR using a theoretical framework formulated at the quark level. A rigorous consequence of this approach is that the quarks (and in principle gluons) -- which are confined inside the bound nucleons -- feel the presence of nearby nucleons and therefore nucleon properties are subtly modified by the nuclear medium~\cite{Guichon:1987jp,Guichon:1995ue}. Our formulation of this approach is based on the Nambu--Jona-Lasinio (NJL) model~\cite{Nambu:1961fr,Nambu:1961tp,Bentz:2001vc} -- which is a chiral effective quark theory of QCD and is readily motivated by QCD's Dyson-Schwinger equations~\cite{Cloet:2013jya}. It provides a natural explanation for the EMC effect~\cite{Cloet:2005rt,Cloet:2006bq,Cloet:2009qs,Cloet:2012td} and has also been used to make predictions for the polarized~\cite{Cloet:2005rt,Cloet:2006bq} and isovector EMC effects~\cite{Cloet:2009qs,Dutta:2010pg,Cloet:2012td}, where the latter provides a significant part of the explanation of the anomalous NuTeV measurement of $\sin^2\theta_W$~\cite{Cloet:2009qs,Bentz:2009yy}.

The nucleon electromagnetic current takes the form
\begin{align}
J^{\mu}(p',p) &= \bar{u}_N(p')\Bigl\{\g^\mu\lf[F_{1p}(q^2)\,\tau_p + F_{1n}(q^2)\,\tau_n\rg] \no \\
&\hs{-6mm}
+ \frac{i\sigma^{\mu\nu}q_\nu}{2\,M_N}\lf[F_{2p}(q^2)\,\tau_p + F_{2n}(q^2)\,\tau_n\rg]\Bigr\}u_N(p),
\label{eq:current}
\end{align}
where $\tau_{p(n)} = \frac{1}{2}\lf(1 \pm \tau_3\rg)$ and all quantities are defined at the same baryon density. For the calculation of the free nucleon form factors we follow Ref.~\cite{Cloet:2014rja} in all respects, including a dressed current with contributions from pions and vector mesons. Form factors for a nucleon in the nuclear medium are determined in a manner analogous to Ref.~\cite{Cloet:2014rja}, except that the quark propagator that enters the Poincar\'e covariant Faddeev equation and the Feynman diagrams that give the nucleon electromagnetic current, takes its in-medium form: $S^{-1}(k) = \sh{k} - M - \sh{V} + i\ve$; where $M$ is the in-medium dressed quark mass and $V^\mu$ the mean vector field, which for the form factor calculation can be eliminated by a shift of the integration variable. The dressed quark mass at a particular density is determined by the minimum of the effective potential for nuclear matter, which is derived from the NJL Lagrangian using hadronization and path integral techniques~\cite{Bentz:2001vc}. Further discussion of this approach can be found in Refs.~\cite{Bentz:2001vc,Mineo:2003vc,Cloet:2009qs}. 

\begin{figure}[tbp]
\centering\includegraphics[width=\columnwidth,clip=true,angle=0]{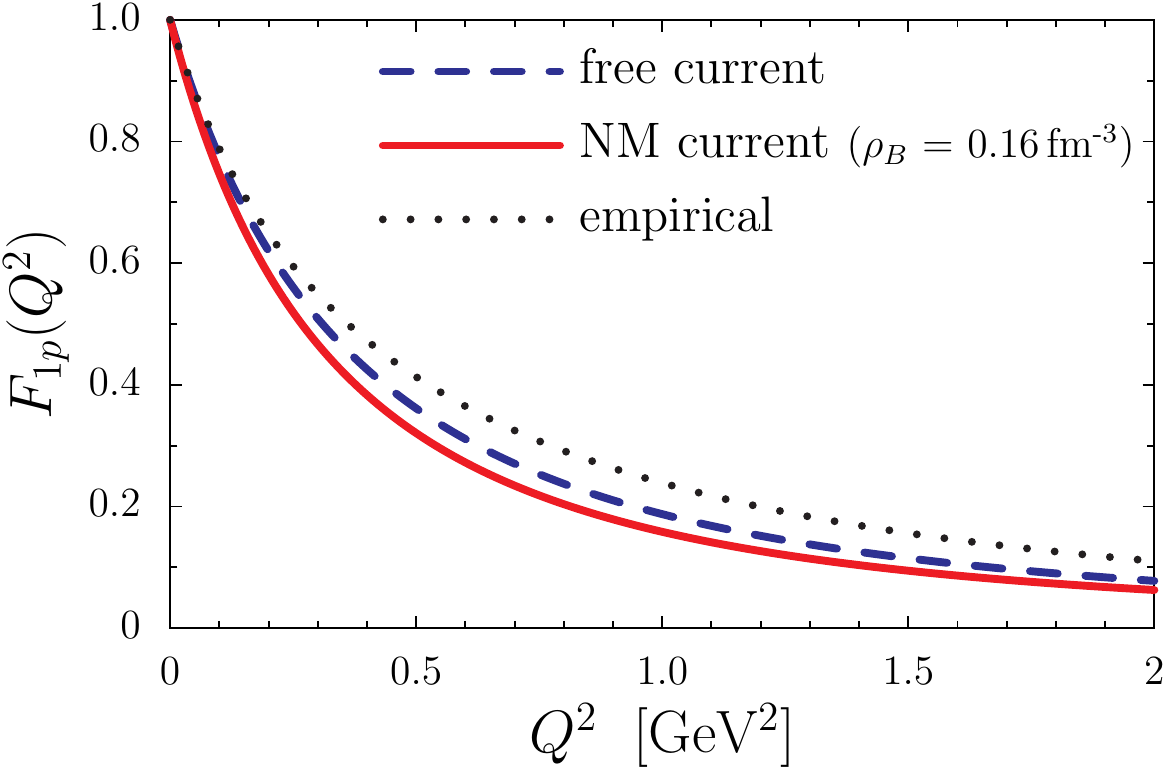}\\[0.8ex]
\centering\includegraphics[width=\columnwidth,clip=true,angle=0]{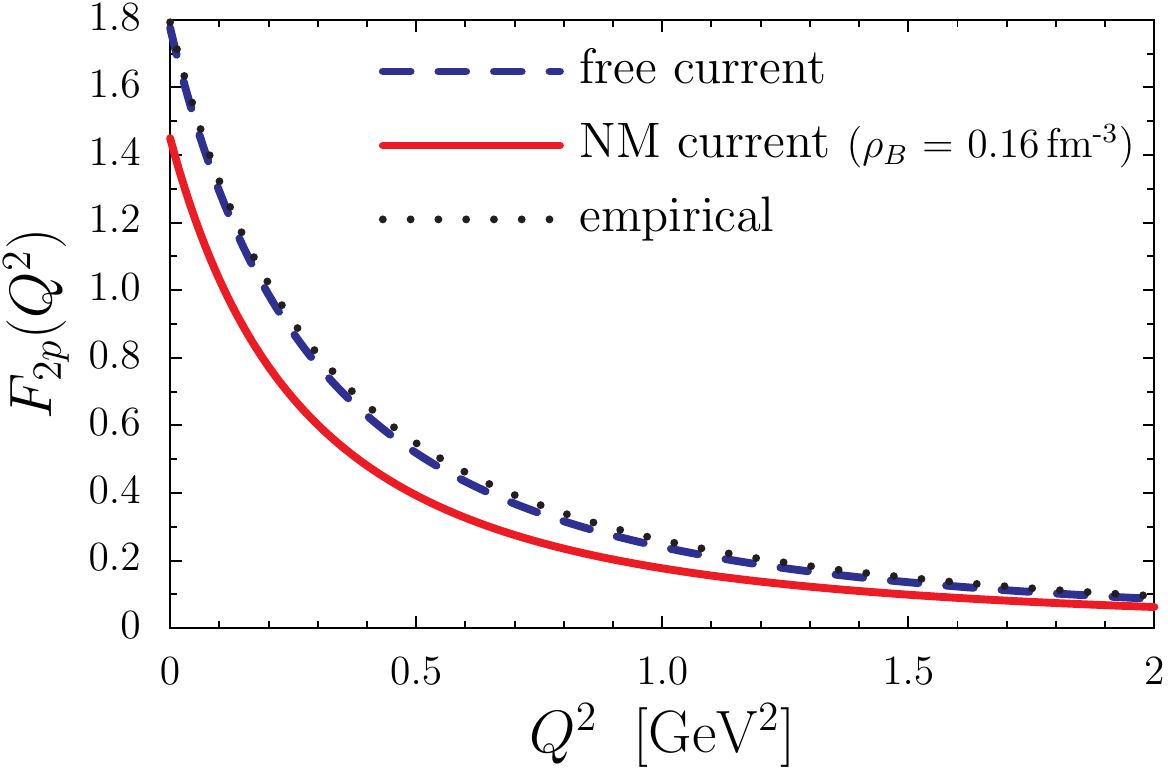}
\caption{(Colour online) Dirac (upper panel) and Pauli (lower panel) form factors
for a free proton and a proton embedded in isospin symmetric nuclear matter (NM current) 
with $\rho_B=0.16\,$fm$^{-3}$. Empirical results are from Ref.~\cite{Kelly:2004hm}.}
\label{fig:formfactors}
\end{figure}

\begin{figure}[tbp]
\centering\includegraphics[width=\columnwidth,clip=true,angle=0]{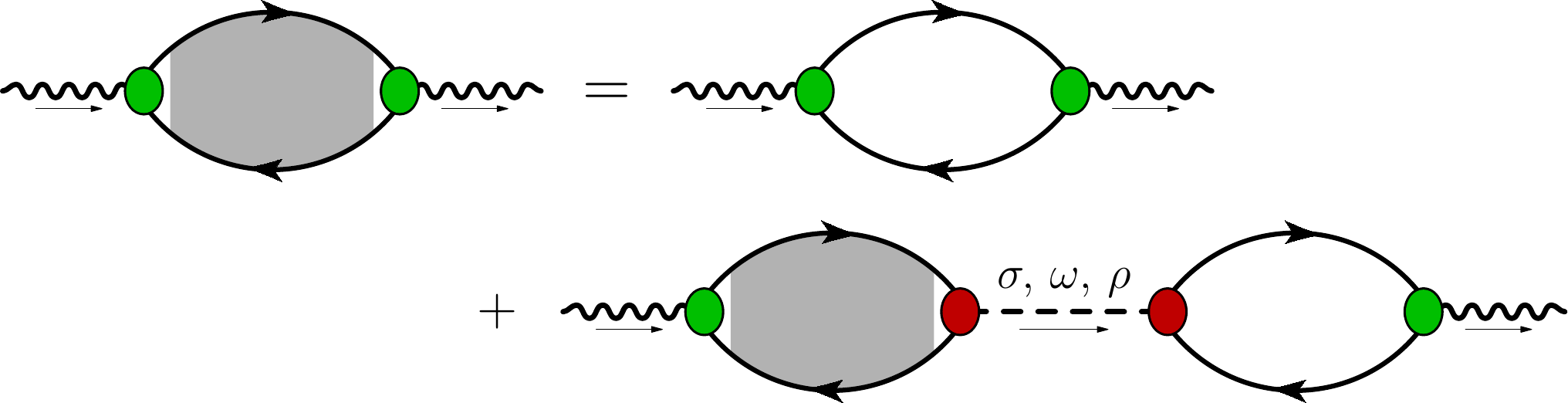}
\caption{(Colour online) Dyson equation for the nucleon polarization (shaded bubble).
For the longitudinal polarization the photon-nucleon vertex
is given by the zeroth component of the current operator given in Eq.~\eqref{eq:current}.}
\label{fig:rpadyson}
\end{figure}

Form factor results for a free proton and a proton embedded in isospin symmetric nuclear matter are illustrated in Fig.~\ref{fig:formfactors}. For the free neutron form factors agreement with experiment is much the same~\cite{Cloet:2014rja} and medium effects are qualitatively similar to that of the proton.\footnote{These results are associated with the current of Eq.~\eqref{eq:current} in the mean field approximation and the explicit density dependent RPA corrections are included via the Dyson equation of Fig.~\ref{fig:rpadyson}.} At nuclear matter saturation density ($\rho_B=0.16\,$fm$^{-3}$) we find that the proton magnetic moment increases by $7\%$, whereas the neutron (anomalous) magnetic moment decreases by $5\%$, relative to their free values. However, the proton anomalous magnetic moment remains almost constant with density; or more generally $F_{2p}(Q^2)/2M_N$ shows only a slight density dependence because changes in $F_{2p}$ are compensated for by those in $M_N$. At saturation density the proton Dirac and charge radii each increase by about $8\%$, whereas the Pauli and magnetic radii grow by half that amount. For the neutron the Pauli and magnetic radii have similar increases as those for the proton, while the charge radius decreases in magnitude by $4\%$ and the Dirac radius more than doubles, reflecting the sensitivity of $F_{1n}$ to small effects. For the Sachs super-ratio relevant to the $(e,e'\,p)$ reactions on $^4$He we find $[G_{Ep}(Q^2)/G_{Mp}(Q^2)]/[G_{E0p}(Q^2)/G_{M0p}(Q^2)] < 1$ (the subscript 0 indicates free proton form factors) with a magnitude consistent with experiment~\cite{Strauch:2002wu,Lu:1997mu}. For the analogous neutron super-ratio we find a result greater than unity, consistent with expectation~\cite{Cloet:2009tx}.

The longitudinal response function can be expressed as the imaginary part of the longitudinal polarization, which in isospin symmetric nuclear matter takes the form~\cite{Wehrberger:1993zu,Horikawa:2005dh}
\begin{align}
R_L(\o,\vect{q}) &= -\frac{2\,Z}{\pi\,\rho_B}\ \text{Im}\,\Pi_L\lf(\o,\vect{q}\rg).
\label{eq:response}
\end{align}
The full result for $\Pi_L\lf(\o,\vect{q}\rg)$ is obtained by solving the Dyson equation illustrated in Fig.~\ref{fig:rpadyson} -- equivalent to the random phase approximation (RPA) -- where the solid lines are the usual finite density nucleon propagators~\cite{Wehrberger:1993zu} and the diagram after the equality is the Hartree result. The nucleon-nucleon interaction, depicted in the third diagram of Fig.~\ref{fig:rpadyson}, is mediated by $\sigma$, $\omega$ and $\rho^0$ exchange and reads
\begin{align}
&V_{NN}(q) = \tau_\sigma(q^2)\lf[\ident\,F_{N}^\sigma(q^2)\rg]_{(1)\cdot(2)} \no \\[0.3em]
&\hs{1mm}+ \tau_\o(q^2)\lf[\gamma^\mu\,3\,F_{1N}^\o(q^2) + \frac{i\sigma^{\mu\nu}q_\nu}{2\,M_N}\,3\,F^\o_{2N}(q^2)\rg]_{(1)\cdot(2)} \no \displaybreak \\
&\hs{1mm}+ \tau_\rho(q^2)\lf[\gamma^\mu\,\tau_3\,F_{1N}^\rho(q^2) + \frac{i\sigma^{\mu\nu}q_\nu}{2\,M_N}\,\tau_3\,F^\rho_{2N}(q^2)\rg]_{(1)\cdot(2)}\!,
\label{eq:nn}
\end{align}
where $[\O]_{(1)\cdot(2)} \equiv \O_{(1)} \cdot \O_{(2)}$ and  the subscripts label nucleon one and two. The quantities $\tau_{\sigma,\omega,\rho}$ are $-i$ times the full $\bar{q}q$ reduced $t$-matrices in the $\sigma$, $\omega$ and $\rho$ channels and represent the propagation of a meson in-medium, including its couplings to the quarks~\cite{Horikawa:2005dh,Cloet:2014rja}. The vector and tensor couplings of the $\omega$ and $\rho$ to the nucleon read
\begin{align}
F_{iN}^{\o(\rho)}(q^2) &= F_{ip}(q^2) \pm F_{in}(q^2),
\label{eq:ffomega}
\end{align}
where $i = 1,\,2$ and for the $\sigma$-nucleon coupling we use
\begin{align}
F_{N}^\sigma(q^2) = g_{\s NN}\lf[F_{1p}(q^2) + F_{1n}(q^2)\rg].
\end{align}
The normalization is rigorously given by $g_{\s NN} = \frac{d\,M_N}{d M}$~\cite{Bentz:2001vc}, which at saturation density takes the value $g_{\s NN} = 1.88$.

For the longitudinal polarization the Dyson equation gives
\begin{align}
\Pi_L\!\lf(\o,\vect{q}\rg) = \Pi^{\g\g}_L - \d\Pi^{\s\o}_L  - \d\Pi^{\rho}_L,
\label{eq:polarization}
\end{align}
where the RPA correlations have the form:
\begin{widetext}
\vspace{-1.3em}
\begin{align}
\d\Pi^{\s\o}_L &= \frac{\tau_\s\lf(1 + \tilde{\tau}_\o\,\Pi^{\o\o}_L\rg) (\Pi^{\g\s}_L)^2
+ \tilde{\tau}_\o\!\lf(1 +  \tau_s\,\Pi^{\s\s}_L\rg) (\Pi^{\g\o}_L)^2 
- 2\,\tau_\s\,\tilde{\tau}_\o\,\Pi^{\o\s}_L\,\Pi^{\g\s}_L\, \Pi^{\g\o}_L}
{\lf(1 + \tilde{\tau}_\o\,\Pi^{\o\o}_L\rg)\lf(1 + \tau_\s\Pi^{\s\s}_L\rg) - \tau_\s\,\tilde{\tau}_\o\,(\Pi^{\o\s}_L)^2},
\qquad
\d\Pi^{\rho}_L = \frac{\tilde{\tau}_\rho\lf(\Pi^{\g\rho}_L\rg)^2}{1 + \tilde{\tau}_\rho\,\Pi^{\rho\rho}_L},
\label{eq:rpapolarization}
\end{align}
\vspace{-0.3em}
\end{widetext}
and $\tilde{\tau}_{\rho(\o)} \equiv q^2/\lf|\vect{q}\rg|^2\tau_{\rho(\o)}$. All polarizations in Eqs.~\eqref{eq:polarization} and \eqref{eq:rpapolarization} are the sum of proton and neutron contributions: $\Pi_L = \Pi_{Lp} + \Pi_{Ln}$; and the Hartree result reads\footnote{This result agrees with Ref.~\cite{Wehrberger:1993zu} but corrects an error in Ref.~\cite{Kurasawa:1985ke}.}
\begin{align}
&\Pi^{\g\g}_{L\l} = F_{1\l}^2\,\Pi_{L\l} + F_{1\l}\,F_{2\l}\,\lf|\vect{q}\rg|^2\,I_{0\l} \no \\
&\hs{6.5mm}
+ \frac{F_{2\l}^2}{8\,M_N^2}\lf[\o^2\tilde{\rho}_{S\l} + 4\,M_N^2\lf|\vect{q}\rg|^2 I_{0\l} + q^2\,I_{2\l}\rg],
\label{eq:hartree}
\end{align}
where $\l = p,\,n$. The polarizations $\Pi^{\g\o}_L$, $\Pi^{\g\rho}_L$, $\Pi^{\o\o}_L$ and $\Pi^{\rho\rho}_L$ are obtained from Eq.~\eqref{eq:hartree} by the appropriate substitutions of Eq.~\eqref{eq:ffomega}. The remaining functions in Eqs.~\eqref{eq:rpapolarization} and \eqref{eq:hartree} are given in appendix A.2 of Ref.~\cite{Wehrberger:1993zu}.

\begin{figure}[tbp]
\centering\includegraphics[width=\columnwidth,clip=true,angle=0]{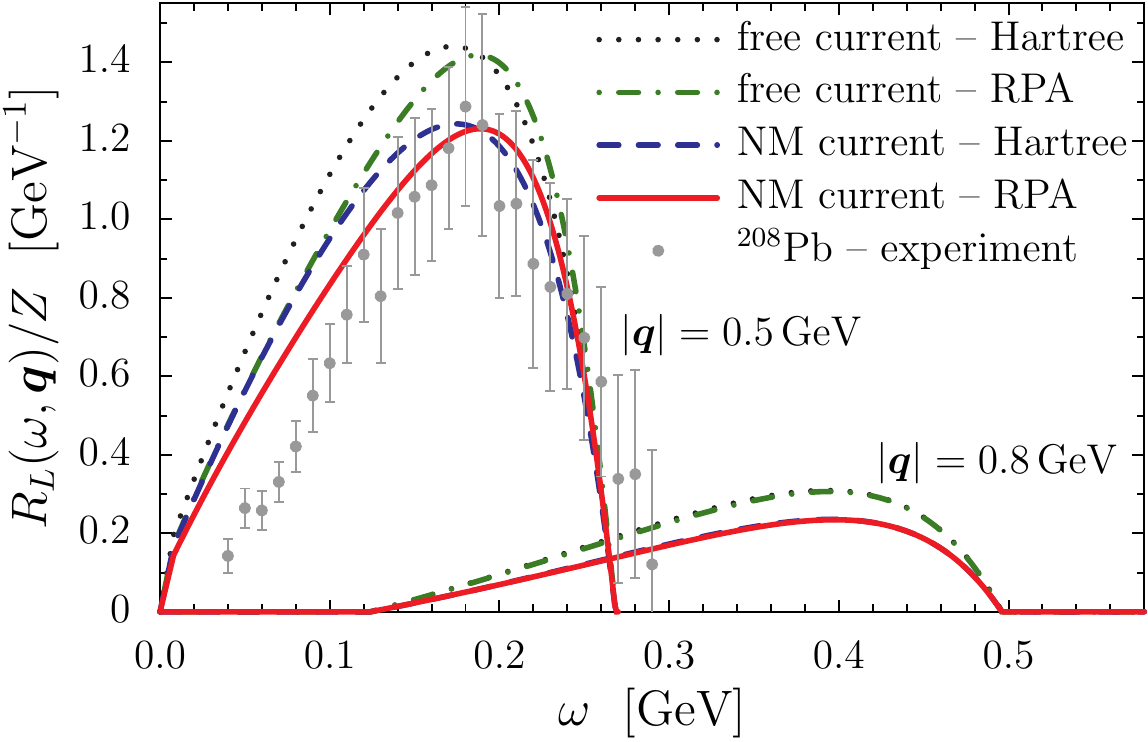}
\caption{(Colour online) Hartree and RPA results for the longitudinal response function in isospin symmetric nuclear matter. Results labelled \textit{free current} are obtained using the free nucleon electromagnetic current operator, whereas the \textit{nuclear matter (NM) current} results use the in-medium current evaluated at $\rho_B = 0.16\,$fm$^{-3}$.  The $^{208}$Pb data at $\lf|\vect{q}\rg| = 0.5\,$GeV is from Refs.~\cite{Zghiche:1993xg,Morgenstern:2001jt}.}
\label{fig:response}
\end{figure}

Hartree and RPA results for the longitudinal response function are given in Fig.~\ref{fig:response} for $\lf|\vect{q}\rg| = 0.5~\text{and}~0.8\,$GeV. The results labelled \textit{nuclear matter (NM) current} are obtained by using the nucleon electromagnetic current operator, meson-nucleon form factors and nucleon propagator evaluated at $\rho_B = 0.16\,$fm$^{-3}$, whereas, for the \textit{free current} results the electromagnetic current operator is that of a free nucleon. For all momentum transfers $\lf|\vect{q}\rg|$, and over all associated energy transfers $\omega$, we find that the longitudinal response function determined with an in-medium nucleon electromagnetic current is quenched relative to the result obtained using a free current. For the Hartree case this result is straightforward to understand from Fig.~\ref{fig:formfactors}, and Eqs.~\eqref{eq:response}, \eqref{eq:polarization} and \eqref{eq:hartree}. For the relevant momentum transfers the proton Dirac form factor contribution to $R_L(\o,\vect{q})$ dominates, and the observed quenching is directly associated with a softer $F_{1p}$ in-medium. For momentum transfers $\lf|\vect{q}\rg| \lesssim 0.7\,$GeV the RPA correlations induce small corrections, shifting the peak in $R_L(\o,\vect{q})$ to slightly larger $\omega$, indicating a net repulsive nucleon--nucleon interaction. For $\lf|\vect{q}\rg| = 0.8\,$GeV correlations play almost no role and $R_L(\o,\vect{q})$ is completely dominated by the Hartree result, that is, $\Pi^{\g\g}_{L}$ of Eq.~\eqref{eq:polarization}. We stress that the longitudinal response function is dominated by the single nucleon contribution~\cite{Lovato:2013cua} and the effect of correlations is relatively small. Our results in Fig.~\ref{fig:response} show good qualitative agreement with the $^{208}$Pb data from Refs.~\cite{Zghiche:1993xg,Morgenstern:2001jt}. In a more sophisticated calculation, including, for example, the neutron excess, Delta baryon and finite nuclear size corrections, the quenching of $R_L(\o,\vect{q})$ would persist and is therefore a robust prediction.

\begin{figure}[tbp]
\centering\includegraphics[width=\columnwidth,clip=true,angle=0]{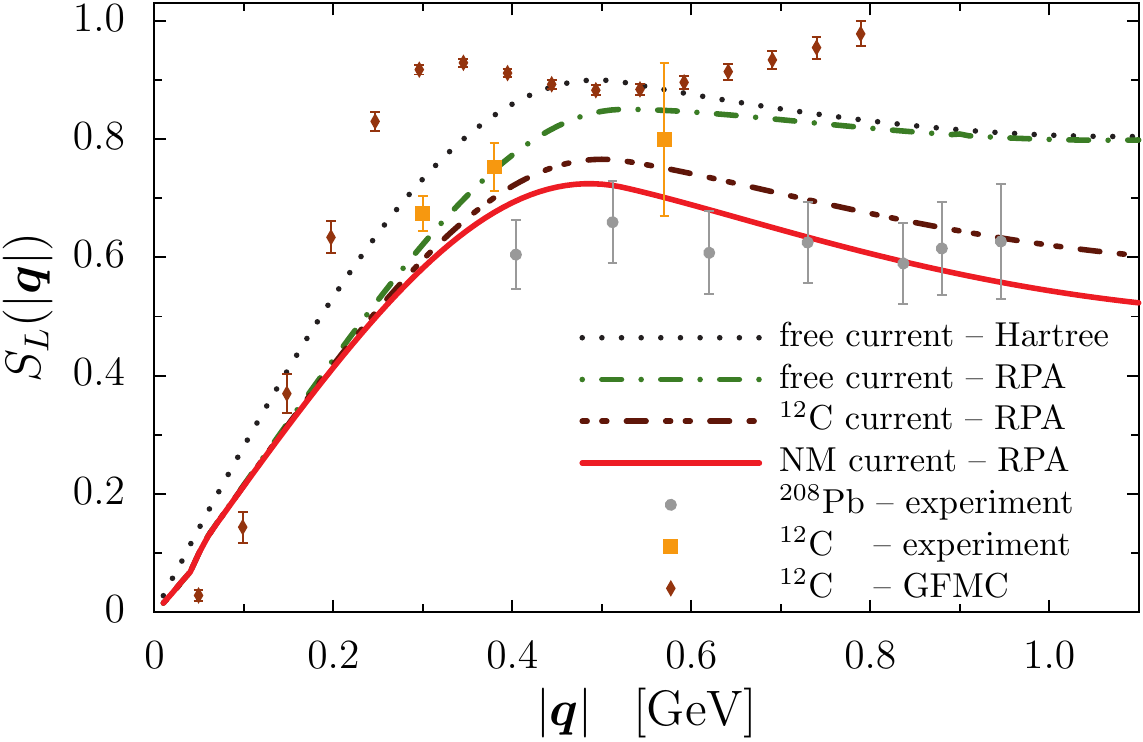}
\caption{(Colour online) CSR determined using a nucleon electromagnetic current operator at: $\rho_B = 0,\,0.1,\,0.16\,$fm$^{-3}$; corresponding to a free nucleon current; at a density typical of $^{12}$C~\cite{Moniz:1971mt};  and at nuclear matter saturation density. The data for $^{208}$Pb is from Refs.~\cite{Zghiche:1993xg,Morgenstern:2001jt} and for $^{12}$C from Refs.~\cite{Barreau:1983ht}, both without the relativistic correction factor of de Forest~\cite{DeForest:1984qe} [to coincide with Eq.~\eqref{eq:csr}].  The GFMC results are taken from Ref.~\cite{Lovato:2013cua}.}
\label{fig:csr}
\end{figure}

Results for the CSR of Eq.~\eqref{eq:csr}, using a nucleon electromagnetic current operator evaluated at three baryon densities:  $\rho_B = 0,\,0.1,\,0.16\,$fm$^{-3}$, are presented in Fig.~\ref{fig:csr}. For the \textit{free current} ($\rho_B = 0$) we illustrate both Hartree and RPA results and find that for $\lf|\vect{q}\rg| \gtrsim 0.7\,$GeV correlations do not materially contribute to the CSR (similar results are found for other baryon densities).  For $\lf|\vect{q}\rg| \simeq 1\,$GeV the CSR for the free  current takes the value $S_L \simeq 0.82$,  which is considerably lower than the non-relativistic expectation of unity.  However, if we take the non-relativistic limit of our result we do recover the naive expectation that the CSR  saturates at unity for $\lf|\vect{q}\rg|$ much greater than the Fermi momentum. Therefore, at $\lf|\vect{q}\rg| \simeq 1\,$GeV we find relativistic corrections to the CSR of roughly $20\%$, which in general are not adequately described by the relativistic correction factor proposed by de Forest~\cite{DeForest:1984qe}. Using the \textit{nuclear matter (NM) current}  ($\rho_B = 0.16\,$fm$^{-3}$) results in a significant further quenching of the CSR for $\lf|\vect{q}\rg| \gtrsim 0.5\,$GeV. For example, at $\lf|\vect{q}\rg| \simeq 1\,$GeV we find that the modification of the nucleon form factors by the nuclear medium results in an additional 30\% reduction in the CSR. The driver of this effect is the medium-induced change to the proton Dirac form factor illustrated in Fig.~\ref{fig:formfactors}. Modification of the Pauli form  factors is less important because their contributions are suppressed by $\lf|\vect{q}\rg|^2\!/4M_N^2$  and $F_{2p(n)}/2M_N$ is largely unchanged in medium. These results demonstrate that the CSR is a particularly sensitive measure of even slight changes in the nucleon form factors, because with increasing $\lf|\vect{q}\rg|$ these effects are cumulative.

This dramatic quenching of the CSR has also been seen in other calculations~\cite{Saito:1999bv,Horikawa:2005dh}, where the internal structural properties of bound nucleons are also self-consistently modified by the nuclear medium. Such observations are consistent with many experiments on various nuclei (e.g. Refs.~\cite{Altemus:1980wt,Meziani:1984is,Deady:1986zz,Baran:1988tw,Chen:1990kq,Zghiche:1993xg,Morgenstern:2001jt}), as illustrated in Fig.~\ref{fig:csr} with a comparison to $^{208}$Pb data from Refs.~\cite{Zghiche:1993xg,Morgenstern:2001jt}. However, calculations that assume an unmodified nucleon electromagnetic current~\cite{DoDang:1987zza,Mihaila:1999nn,Carlson:2001mp,Kim:2006iea}, including the state-of-the-art Green function Monte Carlo (GFMC) result for $^{12}$C from Ref.~\cite{Lovato:2013cua}, consistently find at most modest quenching of the CSR. In Fig.~\ref{fig:csr} the GFMC calculation is contrast with our CSR result, obtained using a nucleon current evaluated at a baryon density typical of $^{12}$C~\cite{Moniz:1971mt}, which again finds a dramatic quenching; the squares are the $^{12}$C data from Refs.~\cite{Barreau:1983ht}, which at the largest $\lf|\vect{q}\rg|$ cannot distinguish between the two results. Both these formalisms have many compelling features. For example,  the GFMC approach has had success at describing properties of $A \leqslant 12$ nuclei~\cite{Wiringa:1994wb,Lovato:2013cua,Carlson:2014vla}, whereas our QCD motivated formalism provides a natural explanation for the EMC effect~\cite{Cloet:2005rt,Cloet:2006bq,Cloet:2009qs,Cloet:2012td}. This impasse over the CSR stands to be resolved however, by forthcoming quasielastic electron scattering results from Jefferson Lab, at high momentum transfer and on a variety of nuclear targets~\cite{E05110}. Verification, or otherwise, of the quenching of the CSR therefore promises to soon reveal critical aspects of the explicit role played by QCD in nuclei.

\begin{acknowledgments}
\textit{Acknowledgements:} This work was supported by the U.S. Department of Energy, Office of Science, Office of Nuclear Physics, contract no. DE-AC02-06CH11357; the Australian Research Council through the ARC Centre of Excellence in Particle Physics at the Terascale and an ARC Australian Laureate Fellowship FL0992247 (AWT); and the Grant in Aid for Scientific Research (Kakenhi) of the Japanese Ministry of Education, Sports, Science and Technology, Project No. 25400270.
\end{acknowledgments}


\end{document}